\documentstyle[11pt,paspconf,epsf]{article}

\newbox\grsign \setbox\grsign=\hbox{$>$}
\newdimen\grdimen \grdimen=\ht\grsign
\newbox\laxbox \newbox\gaxbox
\setbox\gaxbox=\hbox{\raise.5ex\hbox{$>$}\llap
     {\lower.5ex\hbox{$\sim$}}}\ht1=\grdimen\dp1=0pt
     \setbox\laxbox=\hbox{\raise.5ex\hbox{$<$}\llap
	  {\lower.5ex\hbox{$\sim$}}}\ht2=\grdimen\dp2=0pt
	  \def\gax{\mathrel{\copy\gaxbox}}
	  \def\lax{\mathrel{\copy\laxbox}}
	  \def\lta{\lax}
	  \def\gta{\gax}

\begin{document}
\title{Disc instabilities and ``soft" X-ray transients}
\author{J.P. Lasota}
\affil{UPR 176 du CNRS, DARC, Observatoire de Paris, Section de
Meudon, F-92195 Meudon Cedex, France}

\begin{abstract}
I review several problems which arise when one tries to apply the
standard dwarf nova instability model to outbursts of soft X-ray
transients
\end{abstract}


\section{Introduction}

``Soft" X-ray transients (SXTs) are Low--Mass
X-ray Binaries (LMXBs) which appear on the X--ray sky only for a couple of
months. For years or decades they remain in a quiescent state in
which the X--ray emission is just above, or more often below, the
detectability threshold of present X-ray instruments. In LMXBs a
neutron star or a black hole accretes matter lost by a Roche--lobe
filling low--mass companion: a main sequence star (or a star already
significantly 
evolved [King et al. 1996]) or a subgiant. In SXTs neutron
stars are only weakly magnetized so that large scale magnetic fields
play no role in the accretion which proceeds through a disc. The
traditional name of ``soft" X-ray transients may seem misleading
because some of the transient luminosity is also observed in hard X-rays,
but most of the power is emitted in soft (or ultra-soft) X-rays,
unlike the case of ``Hard" X-ray transients which are High Mass
X-ray Binaries (HMXBs) in which matter is accreted on to a strongly
magnetized neutron star.  On the other hand the frequently--used term
``X-ray nova" is even more misleading since it suggests, incorrectly,
that there is similarity between the X-ray transient event and nova
eruptions.

The risetime of SXTs is 2 to 10 days and the outbursts last from 20 to 90
days. Many SXT light--curves can be described as ``FREDs" (fast rise and
exponential decay) but other, more symmetric forms have also been observed
(Chen et al. 1996). Luminosities at outbursts are of the order of
$10^{37}$ -- $10^{38}$ erg s$^{-1}$. The total energy emitted is
typically $10^{43}$ -- $10^{44}$ ergs.

There is no doubt that the outburst is due to an increased accretion
rate in the disc. Until recently two different mechanisms have been
proposed to explain the cause of the transient events: a disc instability,
analogous to the one proposed as the cause of  dwarf nova (DN)
outbursts, and a mass transfer instability in the secondary star that
was supposed to be triggered by X-ray illumination. As was shown by
Gontikakis \& Hameury (1993) (see also Lasota 1996a) the mass transfer
instability model {\sl cannot} account for the observed properties
of SXTs, so that it should no longer be quoted as a possible model
for SXT outbursts.

It is clear however that the DN disc instability model (DIM) cannot be
applied to SXTs without modifications (Lasota 1996a,b; Cannizzo et al
1996; Mineshige 1996; Lasota et al. 1996; van Paradijs 1996). One
should also stress that while X-ray illumination of the secondary
cannot trigger the outburst, irradiation of the secondary and of the
disc by X--rays during the outburst may play an important role in
the SXT phenomenon (Chen et al. 1994; Augusteijn et al 1994).

In this article I review possible modifications of the DIM that would
allow one to apply it successfully to SXT outbursts. In particular I
discuss the model in which the inner regions of the accretion flow are
advection dominated (Narayan et al. 1996) and the role played in the
model by X--ray illumination of the disc (van Paradijs 1996).

\section{The `standard', dwarf--nova disc instability model}

The cause of the disc instability which operates in DNs is due to the
partial ionization of hydrogen at $\sim 10^4$ K. The resulting abrupt
change in opacities between 6000 and 10000 K makes the disc (locally)
thermally and viscously unstable. For accretion rates that correspond
to this range of temperatures a stable equilibrium is impossible and
one expects a limit--cycle behaviour. To obtain a global
instability that would correspond to the observed amplitudes and
durations of DN outbursts one has to assume that the viscosity
$\alpha$ parameter depends on the local disc properties in such a way
that it is 4 to 10 times larger in outburst than in quiescence. In the
`standard' disc instability model one assumes that the mass transfer
from the secondary is constant during the outburst cycle.

The thermal instability model, however, cannot by itself account for
the so called `super--outbursts' observed in SU UMa systems. In this
case an additional mechanism is needed to account for the higher
amplitude and longer duration of the outbursts. Osaki (see this volume)
has proposed a thermal--tidal disc instability model in which tidal
forces increase the effective viscosity when the disc's outer radius is
larger that a certain critical value. This model can work only for mass
ratios (secondary/primary) smaller than $\lta 0.25$ and cannot
therefore explain the high amplitude and very long outburst observed in
U Gem (we do not call this unusual outburst a `superoutburst', since no
`superhump' has been observed [and was not supposed to be observed; see
Osaki in this volume], but its light--curve looks pretty much like a
usual `superoutburst').

In the case of systems in which only superoutbursts are observed
(unlike the usual SU UMa stars, where superoutbursts are separated by
cycles of `normal' outbursts) the standard and the thermal--tidal disc
instability models can be applied more or less successfully only if an
additional assumption is made: the value of $\alpha$ in quiescence must
be very small: $\sim 10^{-4} - 10^{-5}$ (Smak 1993) instead of 
$\sim 10^{-2}$ in most of the DN systems. The physical reason for such
low $\alpha$ values remains however unexplained.

The standard DIM also cannot account for the properties of many
quiescent DN systems. According to DIM, in quiescence, {\sl the whole}
accretion disc has to sit on the cold branch of stable disc equilibria.
This implies that the surface density $\Sigma$ of the cold disc must be
lower than the value $\Sigma_{\rm max}$ corresponding to the maximum
allowed effective temperature of a cold stable disc.  The maximum
surface density of the cold disc is approximated by 
\begin{equation}
\Sigma_{\rm max} \approx 672.8 \ \alpha_{0.01}^{-0.80} r_{10}^{1.10}
						    m_{1}^{-0.37}
						    \label{sigmax}
\end{equation} the corresponding critical accretion rate,
above which the disc is unstable, can be approximated as
\begin{equation} \dot M_{\rm crit,C} \approx 4.7 \times 10^{15}\
                       r_{10}^{2.65}  m_1^{-0.88}\ {\rm g\ s^{-1}}
			      \label{dotmax}
\end{equation} 
(Hameury et al. 1996), where $\alpha = 0.01
\alpha_{0.01}$, $m_1$ is the white dwarf mass in solar units and
$r=10^{10} r_{10}$ cm is the radius in the disc.

In quiescence the local value of the accretion rate has to be everywhere
smaller than $\dot M_{\rm crit, C}\left(r \right)$, and the quiescent DN disc
is not in equilibrium since $\dot M \neq {\rm constant}$.

Eq. (\ref{dotmax}) implies that for $m_1=1$ and a white dwarf radius
$r_{10} =.05$ the accretion rate at the inner disc edge should be less
than $\sim 6 \times 10^{12} \ {\rm g \ s^{-1}}$.  X-ray observations of
quiescent DNs, however imply values of the accretion rate on to the
white dwarf of $\sim 10^{14} - 10^{15}\ {\rm g\
s^{-1}}$ (Erakleous et al. 1991; Mukai \& Shiokawa 1993), if the X-rays
are emitted by the accretion flow, as seems indeed to be the case,
since the secondaries are unlikely to emit the observed luminosity. As
will be discussed below a similar problem arises for SXTs (Lasota
1996a,b; Cannizzo et al. 1995).

The standard DIM cannot therefore account for the properties of the
inner accretion flow: according to the observations, the cold quiescent
accretion disc cannot extend down to the surface of the white dwarf. A
boundary layer emission is of no help since the required {\sl accretion
rate} on to the central object would be too high to satisfy the
requirement of the DIM.

Since $\dot M_{\rm crit, C}$ increases with radius while the accretion
luminosity decreases, the difficulty can be solved by increasing the
inner disc radius. Two ways of achieving this have been proposed.  One
(Livio \& Pringle 1992) is to assume that the white dwarf possesses a weak
magnetic field which is strong enough to disrupt the inner disc in
quiescence. In the other, at low accretion rates
(low densities) the inner disc will evaporate, forming an extended
coronal flow with a low radiative efficiency. A similar idea in the
case of SXTs was proposed by Narayan et al. (1996). This advection--dominated 
accretion flow model (ADAF) will be discussed below. Of course in the case
of a black hole the magnetic field hypothesis cannot work.

It is also interesting to note that `removing' the inner disc regions
may solve the so--called ``optical -- UV  delay problem"  in modeling
outbursts of some DNs. In such DNs one observes a substantial delay in
the optical and UV outbursts: the optical event begins first, and the UV
one follows only after several hours. In the `standard' DIM model the
heating front propagates rather quite quickly from the outer (optical emission)
disc towards the inner (UV emitting) disc regions and models cannot
account for the observed delay. In a disc in which the inner regions have
either evaporated or been disrupted by a magnetic field, the heating front
will be stopped at the disc's inner edge, at which UV emission is weak.
The inner disc reaching to the white dwarf surface will be rebuilt in 
viscous time, thus reproducing the observed delay (see Livio and
Pringle 1992).

As has been mentioned above, if WZ Sge type outbursts were to be
explained by the DIM one would have to assume an $\alpha$ in quiescence
much smaller than in other DN outbursts. As shown by Hameury et al. in
these proceedings an alternative model may explain WZ Sge properties
without assuming unusually low $\alpha$ values. Also in this model one
requires a truncated inner disc (see also Lasota et al. 1995).
It is worth noting that WZ Sge systems are very similar to SXTs
(Kuulkers et al. 1996; Lasota 1996a,b). 

One can conclude therefore that in many cases the DIM has to be
modified to be able to account for the properties of DN outbursts. 
The required modifications concern the inner disc regions. 
Similar problems are encountered when the DIM is applied to
SXT outbursts.

\begin{figure}
\plotone{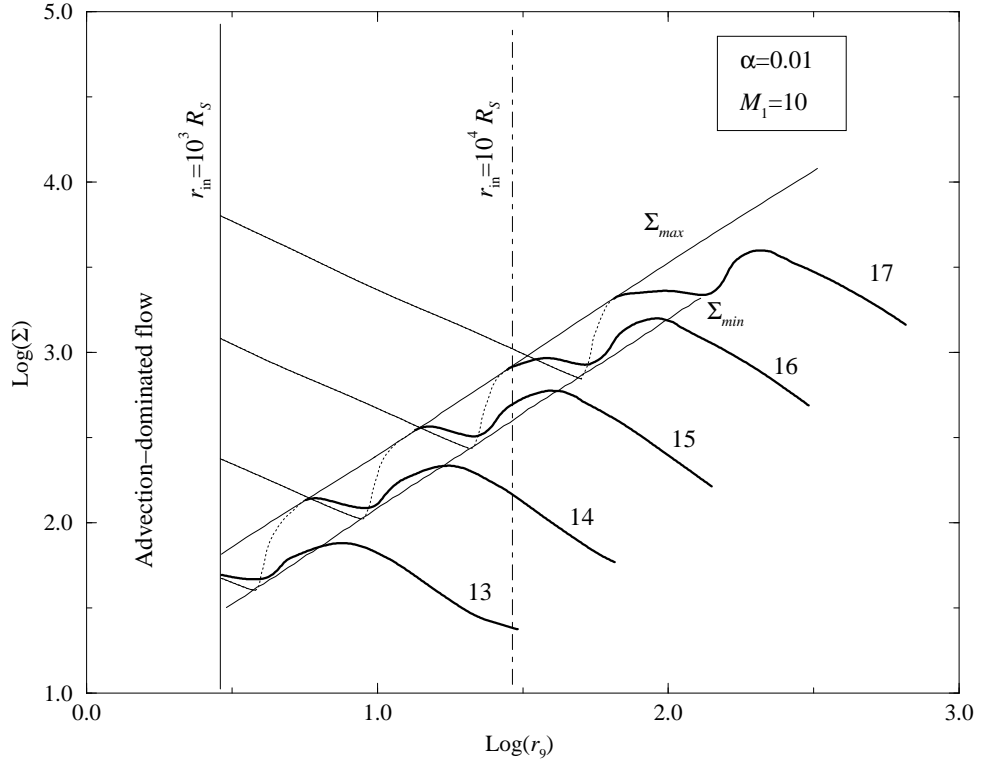}
\caption{The surface density profiles of equilibrium accretion
disc configurations around a black hole of $10M_{\odot}$ with $\alpha=0.01$. 
Cold, stable equilibria are represented by solid lines, unstable
by dotted lines. The hot stable solutions correspond
to the straight lines terminating on the $\Sigma_{\rm min}$ line.
The inner regions
of thin discs could be truncated at $\sim 10^{3-4}$ Schwarzschild
radii ($R_s$).}
\end{figure}
\section{The disc instability model for SXTs}

\subsection{The stability criterion}

At effective temperatures $T_{\rm eff} > T_{\rm crit} \approx 6500$ K
an accretion disc is thermally stable (on the `hot' branch). Since the
effective temperature for a stationary accretion disc is given by
\begin{equation} T_{\rm eff}= \left( \frac{3GM\dot M}{8\pi\sigma
r^3}\right)^{1/4} \label{teff} \end{equation} where $\sigma$ is the
Stefan-Boltzmann constant, it is enough for stability that $T_{\rm eff}
> T_{\rm crit}$ at the outer disc radius. On Fig. 1 the hot stable
solutions are represented by straight lines terminating on the
$\Sigma_{\rm min}$ line, where $\Sigma_{\rm min}$ corresponds to the
minimum surface density of the hot solution. From Eq. (\ref{teff}) and
the relation between the disc radius and the orbital period one can
obtain an expression for the critical {\sl mass transfer rate} above
which, for a given orbital period, the disc is stable (Smak 1983; see
also King et al. 1996): 
\begin{equation} \dot M_{\rm crit, H} \approx
1.8 \times 10^{17} P_3^2\ {\rm g \ s^{-1}} \label{dotmdn}
\end{equation} 
where $P_3= P/(3$hr). In deriving Eq. (\ref{dotmdn}) the
relation 
\begin{equation} r_{\rm D} \approx 3.0 \times 10^{10} m_1^{/3}
P_3^{2/3} {\rm cm} \label{dradius} \end{equation}
has been used, where
$r_{\rm D}$ is the outer disc radius.

Figure 1 illustrates the significance of the stability criterion given
by Eq. (\ref{dotmdn}). One can see, for example, that the straight line
part of the equilibrium curve for $\dot M = 10^{17}$ ends on the
$\Sigma_{\rm min}$ line at a radius $\sim 6.5 \times 10^{10}$ cm, which,
according to Eq. (\ref{dradius}), corresponds to $P \approx 3$ h, in
agreement with Eq. (\ref{dotmdn}).

The criterion $\dot M > \dot M_{\rm crit, H}$ is a sufficient condition
for disc stability (or its negation a necessary condition for instability).
It is not a necessary condition for stability since a disc for which
$\dot M < \dot M_{\rm crit, C}\left(r_{\rm in}\right)$ (see Eq. \ref{dotmax}) 
is globally stable since it is everywhere on the `cold' stable branch.
If one assumes, as does the standard version of the DIM, that
the disc extends down to the white dwarf surface, mass transfer rates
are considered to be too low ($\dot M \lta 10^{12} - 10^{13}$ g s$^{-1}$)
to be of practical interest and  $\dot M > 
\dot M_{\rm crit, DN}$ is assumed to be a necessary and sufficient 
stability condition.

In the case of some dwarf novae (as discussed above) and for all
black--hole SXTs (BHSXTs) the inner accretion disc has to be truncated
(as we will show below) so the condition $\dot M > \dot M_{\rm crit,
H}$ is not a necessary condition for stability even in practical
applications. For WZ Sge Lasota et al. (1995) considered a model in
which the inner disc is truncated at $r_{\rm tr} \approx 2.5 \times
10^9$ cm (for $m_1=0.4$) and thus marginally stable (see also Warner et
al. 1996 and Hameury et al. in these proceedings).  Narayan et al.
(1996)  proposed a quiescent SXT model in which the outer dwarf--nova
type disc is truncated at $10^{3} - 10^{4} r_{\rm S}$, where
$r_{\rm S} = 2GM/c^2 r$ is the Schwarzschild radius, so that this outer
disc could be globally stable for mass--transfer rates $\dot M \sim
10^{14} - 10 ^{16}$ g s$^{-1}$ (see e.g. Lasota et al. 1996 and Narayan
in these proceedings). These models  will be discussed in the next
section. Before that, in the next subsection, we will discuss how X--ray
illumination may affect the stability criterion.

\subsection{The effect of X--ray illumination on the disc stability}

The stability criterion requiring mass transfer rates to be larger than
the value given by Eq. (\ref{dotmdn}) is derived from
the requirement that the disc's effective temperature be higher 
than $\sim 6500$ K everywhere. In the standard DIM it is assumed that
the disc temperature is determined by power released by viscosity
(see Eq. (\ref{teff})).

\begin{figure}
\plotone{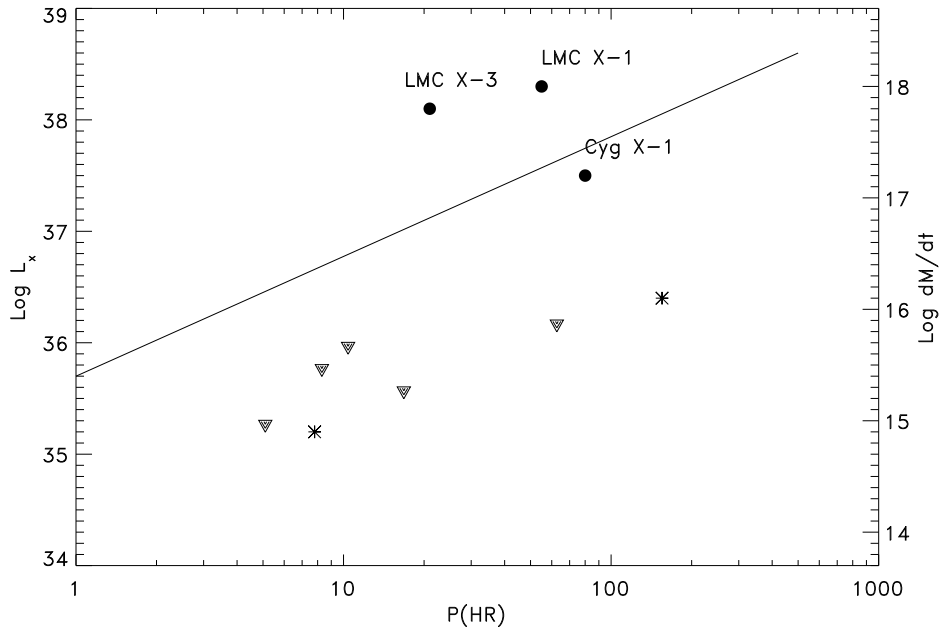}
\caption{X-ray luminosity (and average mass transfer rate)  as
a function of orbital period for persistent and  transient LMXB with
black holes. The transients with known recurrence times (A0620--00 and
V404~Cyg) have been
indicated with an asterisk, the other transients with a triangle.  The
straight line indicates the separation between persistent and 
transient sources derived here for black holes of 10 M$_{\odot}$ (Eq.
\ref{dotirr}). The figure also includes (indicated by dots) the three persistent
high-mass X--ray binaries Cyg X--1, LMC X--1 and LMC~X--~3, at the fiducial
positions that they would have occupied if they had been LMXB
with an equally large accretion disc, i.e.,  Roche lobe of the X-ray
source; (from van Paradijs 1996).}
\end{figure}

Recently van Paradijs (1996) pointed out that since 
accretion discs in LMXBs
are likely to be strongly affected by X-ray illumination from the central
source the heat released by the X-ray  irradiation should
stabilize the disc, so that LMXB accretion discs would be stable at 
accretion rates
lower than those given by Eq. (\ref{dotmdn}). By taking  X--ray heating
into account, van Paradijs (1996) obtained a new
critical mass transfer rate above which accretion discs in LMXB are 
globally stable.
This critical rate can be written as (King et al. 1996):
\begin{equation}
\dot M_{\rm crit, irr } \approx 3.2 \times 10^{15} m_1^{2/3}P_3^{4/3}\
 {\rm g\  s^{-1}}
\label{dotirr}
\end{equation}
The new critical $\dot M$ is represented on Fig. 2. on which the parameters of 
black hole X-ray binaries are plotted. The new criterion seems to separate
persistent sources from transient ones (or variable like Cyg X-1). 
The same is true for LMXBs containing neutron stars (van Paradijs 1996).

Like Smak's criterion for DNs, the van Paradijs criterion gives a 
sufficient condition for stability: it states
that a stationary disc for which $\dot M > \dot M_{\rm crit, irr }$
is globally stable. Unlike Smak's criterion, the van Paradijs
condition is non--local: the X--ray illumination of the outer disc
depends on the accretion rate on to the central object and on the shape
of the disc, which must allow the outer disc ``to see" the central
X--ray source. A stationary disc is concave (in the range of parameters
of interest) so the outer disc is exposed to the irradiation by the central
source. 

This is not the case of non--stationary quiescent discs according to
the DIM. There, because of the convex disc shape, its outer regions are
shielded from the central X-ray source (Cannizzo 1994).  So, do
transient sources obey the criterion $\dot M < \dot M_{\rm crit, irr}$
? This will depend on their history, how they have arrived at the
present level of mass transfer rate. If the present mass transfer rate
was achieved by going through a sequence of  mass transfer rates higher
than the present one, the van Paradijs criterion will be obeyed. If,
however, the present state was achieved from lower mass transfer rates,
i.e.  through a sequence corresponding to globally unstable disc
configurations, the mass transfer rates could be higher than the
critical value given by Eq. (\ref{dotirr}) because the outer disc
would never be exposed to the central X--ray source (Cannizzo 1994).

A quick look at Figure 2. shows, however, that (estimated) mass transfer
rates for BHSXTs are in fact well below the limit given by  Eq.
(\ref{dotirr}). 
On the other hand, all persistent sources are above the
van Paradijs limit 
and no SXT has a mass transfer rate superior to
$\dot M_{\rm crit, irr}$, confirming that this critical value provides a
sufficient condition for disc stability.  This does not mean, however, as
we will see below, that accretion discs in BHSXT {\sl are unstable}.

\subsection{The standard DIM applied to SXTs}

The application of the standard DIM to SXT outbursts (Mineshige \&
Wheeler 1989; Cannizzo et al. 1995) leads to the same difficulties
as those encountered in the case of X--ray emitting quiescent dwarf
novae (see section 2) but these difficulties are more acute. The
DIM, in its standard version, requires  the whole disc to be in the
cold, stable equilibrium. Since the critical accretion rate (Eq.
\ref{dotmax}) scales as $r^{2.5}$ and a black hole (or neutron star)
radius is about 10$^3$ times smaller than a white dwarf radius, the
quiescent accretion rates on to the central body required by the DIM are,
in the case of SXTs, ridiculously small ($\lta 10^6$ g s$^{-1}$) and in
contradiction with observations.

Indeed, several BSXTs have been observed by GINGA (Mineshige et al. 1992)
and ROSAT (see Verbunt 1996), and two systems, A0620-00 and V404 Cyg,
have been detected at levels corresponding to accretion rates (assuming
an efficiency of 0.1, see below) of at least $\sim 1.3 \times
10^{11}$~g~s$^{-1}$ for A0620-00 and $3 \times 10^{12}$~g~s$^{-1}$ for
V404 Cyg (Mineshige et al. 1992; McClintock, et al. 1995;
Verbunt 1996).  These accretion rates are several orders of magnitude
larger than the values of $\lta 10^6$ g s$^{-1}$ required by the
standard DIM.  

We note that in the case of A0620--00, the prototypical BSXT, the
optical and UV luminosities suggest a mass transfer rate of $\sim 6
\times 10^{15}$~g~s$^{-1}$ (McClintock et al. 1995).

One could argue that since the observed X--ray fluxes are low (in the
case of A0620-00 only 39 counts were detected) there is no
reason to assume that these X-rays are emitted by the accretion flow.
One can, however, exclude (Verbunt 1996) that  X-ray
emission from the companion (except for A0620-00 for which the evidence
is marginal). Future observations will have to decide on the origin of
the X--rays observed in quiescent SXTs (V404 Cyg is here the most
promising system). However, as mentioned above (see Cannizzo et al.
1995), the problem of inconsistency between requirements of the DIM and
observations is analogous to the one encountered for DNs so we would
find it surprisising if the X-rays were not emitted by matter accreting
on to the central object.

\section{Advection dominated accretion flows in SXTs}

Let us forget for the moment about the requirements of the DIM and
consider the implications of the X--ray and optical/UV observations for
the structure of quiescent accretion flows in SXTs. If we assume that
the accretion flow forms a Shakura--Sunyaev disc down to the last stable
orbit around the black hole, the efficiency of accretion is (in the
Newtonian approximation, which is sufficient for the order of magnitude
arguments) $\eta \sim GM/rc^2$, i.e $\sim 0.1$ for a
black hole. One can then deduce accretion rates from luminosities by
using the relation $\dot M = \eta L / c^2$. In such a `standard' framework
the mass transfer rate in A0620-00 is $6 \times 10^{15}$ g s$^{-1}$
whereas the
accretion rate into the black hole is $\sim 3 \times 10^{11}$ g s$^{-1}$
(McClintock et al. 1995). So there is a four orders of magnitudes
difference between the rate in which matter is deposited at the outer
disc edge and the rate at which it is lost in the black hole. The large
differences between mass transfer rates and accretion rates on to the
central accreting object are typical of quiescent, nonstationary
discs of the DIM, which leads McClintock et al. (1995) to the conclusion
that their observations of A0620-00 {\sl confirm} the DIM. As was
mentioned in the previous section however, the DIM requires a 10 orders
of magnitude difference, so that in fact McClintock et al.'s (1995)
observations refute the DIM for SXTs. 

The difficulty in modelling quiescent SXT discs is due to the
discrepancy between the `observed' mass transfer rate and the accretion
rate on to the central object deduced by assuming a radiative
efficiency of 0.1.  This discrepancy cannot be explained in the
framework of the DIM.  There are other systems in the Universe from which
such discrepancy is inferred: our Galactic Center, nuclei of giant
elliptical galaxies and some weak active galactic nuclei. As this was
first realized by Rees (1982), in such systems the efficiency for
conversion of rest mass into luminosity may be rather low, i.e. 
much lower than 0.1, if only a small fraction of the heat released by
accretion can be radiated away during the infall of the material on to
the black hole. In such {\sl advection--dominated accretion flows}
(ADAFs) the accretion rate on to the black hole is obviously much
higher than the one deduced by using a 0.1 efficiency and in particular
the accretion rate can be constant in the flow. Fabian \& Rees (1995) have
suggested that accretion flows in the nuclei of giant elliptical galaxies
are advection--dominated.

Recent work on optically thin ADAFs (see Narayan in these proceedings)
provided reliable models that can be used to describe various
systems.

The two-temperature model of Narayan \& Yi (1995) has been applied to
to the Galactic Center (Narayan et al. 1995), and the LINER NGC 4258 
(Lasota et al. 1996)[in this case, there are, model--dependent,
arguments against
this object containing an ADAF -- see articles by Maloney and by 
Begelman in these proceedings]. Narayan et al. (1996) proposed a model
of quiescent SXTs in which an outer cold Keplerian disc has its inner edge
at a large transition radius ($10^3 - 10^4 r_S$) at which the flow becomes
very hot and advection dominated. Mineshige (1996) used such a framework
to discuss the mechanism driving SXT outbursts.

In the Narayan et al. (1996) model the accretion rate is constant
through the disc. The X-rays are emitted by the ADAF while the
UV/optical radiation originates in the outer cold disc.  As was shown
by Lasota et al. (1996), however, the UV/optical flux cannot be emitted
by this part of the disc. This is a problem for {\sl all} models.  In
the new version of this model presented at this meeting by Narayan,
most of the observed radiation is emitted by the ADAF with a transition
radius at about 6000 $R_S$ for A0620-00 and 25000 $R_S$ for V404 Cyg
(Narayan et al. 1997)
but the value of transition radius is not very well constrained by the
models.  The UV flux is due to the synchrotron emission of the ADAF.
The outer disc is stationary.

If the outer disc is indeed stationary, what is the origin of the outbursts?
Kuulkers et al. (1996) claim that the outburst mechanism must be the same
both in WZ Sge systems and in SXTs. I agree with them, but disagree about the
cause. As in the Hameury et al. model of WZ Sge, the outbursts
of SXTs could be due to an enhanced mass transfer rate
(see e.g. Lasota 1996b) which brings the
marginally stable disc into an unstable regime. It is interesting to note
that the positions of the BHSXTs in Fig. 2. correspond to marginally
stable discs truncated at $r_{\rm tr}\approx 10^3 - 10^4 r_S $, i.e.
values required, by the Narayan et al. model. As can be seen from Fig. 1
$r_{\rm tr}\approx 10^4 r_S $ corresponds to a stable outer disc for 
$\dot M \gta 10^{16}$ g s$^{-1}$ and for $\dot M \gta 10^{15}$ g s$^{-1}$
the outer disc is stable for $r_{\rm tr}\gta 3 \times 10^3 r_S $.

\acknowledgements

I am gratful to John Cannizzo, Jean--Marie Hameury, Andrew King,
Ramesh Narayan and Craig Wheeler for enlightening discussions on
the subject of SXTs.

\vfill\eject

\section{Discussion}

{\sl Marina Romanova}: Some X--ray Novae show jets. You did not mention 
 this fact in your talk. Are there any models which may explain jets? There
are models where disc may be magnetically unstable and outflows are possible.\\
\medskip

\noindent {\sl J.P. Lasota}: I did not mention jets from SXTs because I don't 
like them.
I am full of admiration for the observations of Felix Mirabel and Bob Hjellming,
but I wish they didn't find jets. I have no explanation for them and I 
believe that what you call `models' are no more than `scenarios'. The possible
connection between ADAFs and jets has been discussed by Ramesh Narayan at this 
workshop.

\end{document}